\documentstyle[epsfig]{aipproc}

\begin{document} 

{\footnotesize \it \noindent 
   Proc.\ 5th Compton Symposium (Portsmouth NH, September 1999)}

\title{Evidence for a discrete source contribution to low-energy 
continuum Galactic $\gamma$-rays}

\author{Andrew W. Strong$^*$ and Igor V. Moskalenko$^{*\dagger\ddagger}$}
\address{
$^*$Max-Planck Institut f\"ur extraterrestrische Physik, 85740
Garching, Germany\\ 
$^{\dagger}$ Institute for
Nuclear Physics, M.V.  Lomonosov Moscow State University, Moscow, Russia\\
$^{\ddagger}$ LHEA NASA/GSFC Code 660, Greenbelt, MD 20771, USA
}

\maketitle

\begin{abstract} 
Models for the diffuse Galactic continuum emission and
synchrotron radiation show that it is difficult to reproduce observations of
both of these from the same population of cosmic-ray electrons.  This indicates
that an important contributor to the emission below 10 MeV could be an
unresolved point-source population.  We suggest that these could be Crab-like
sources in the inner Galaxy.  Alternatively a sharp upturn in the electron
spectrum below 200 MeV is required.

\end{abstract}

\section{Introduction}

Although `diffuse' emission dominates the COMPTEL all-sky maps in the
energy range 1--30 MeV, its origin is not yet firmly established; in
fact it is not even clear whether it is truly diffuse in nature.  This
is in contrast to the situation at higher energies where the close
correlation of the EGRET maps with HI and CO surveys establishes a
major component as cosmic-ray interactions with interstellar gas.
This paper discusses recent studies of the low-energy diffuse
continuum emission based on the modelling approach described in
\cite{sm98}.  The high energy ($>$ 1 GeV) situation is addressed in
\cite{smr98,Strong2000}.  The present work uses observational results
reported in \cite{Strong98}; new imaging and spectral results from
COMPTEL are presented in \cite{Bloemen2000} but differences are not
important for our conclusions.

\begin{figure}[h] 
   \centerline{\epsfig{file=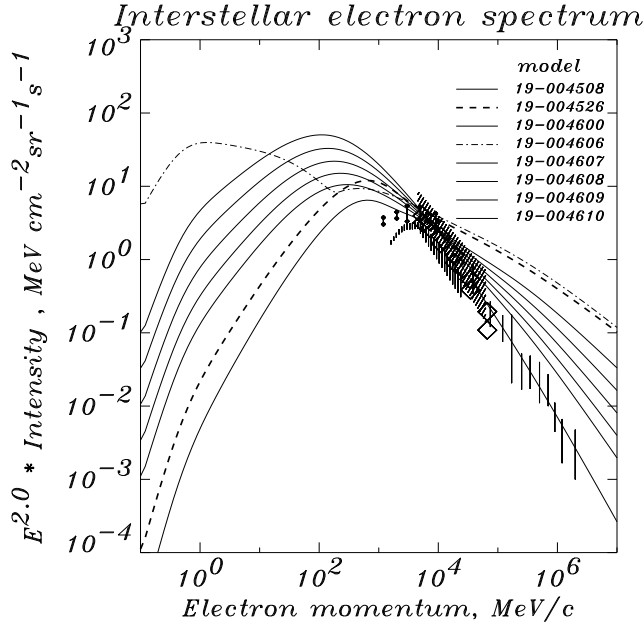,height=3.4in,width=3.5in}}
\vspace{10pt}
\caption{Electron spectrum after propagation for   various electron
injection  spectra. Injection index 2.0 to 2.4 (narrow full
lines). Also shown is a  spectrum which reproduces the high-energy
$\gamma$-ray excess (dashed line) and  a spectrum with a sharp  upturn
below 200 MeV which can reproduce the low-energy  $\gamma$-rays
without violating synchrotron constraints (dash-dot line). The  thick
solid line is a spectrum consistent with both local measurements and
synchrotron constraints. For the data compilation see
\protect\cite{smr98}.       
}
\label{electrons}
\end{figure}

\begin{figure}[t] 
   \epsfig{file= 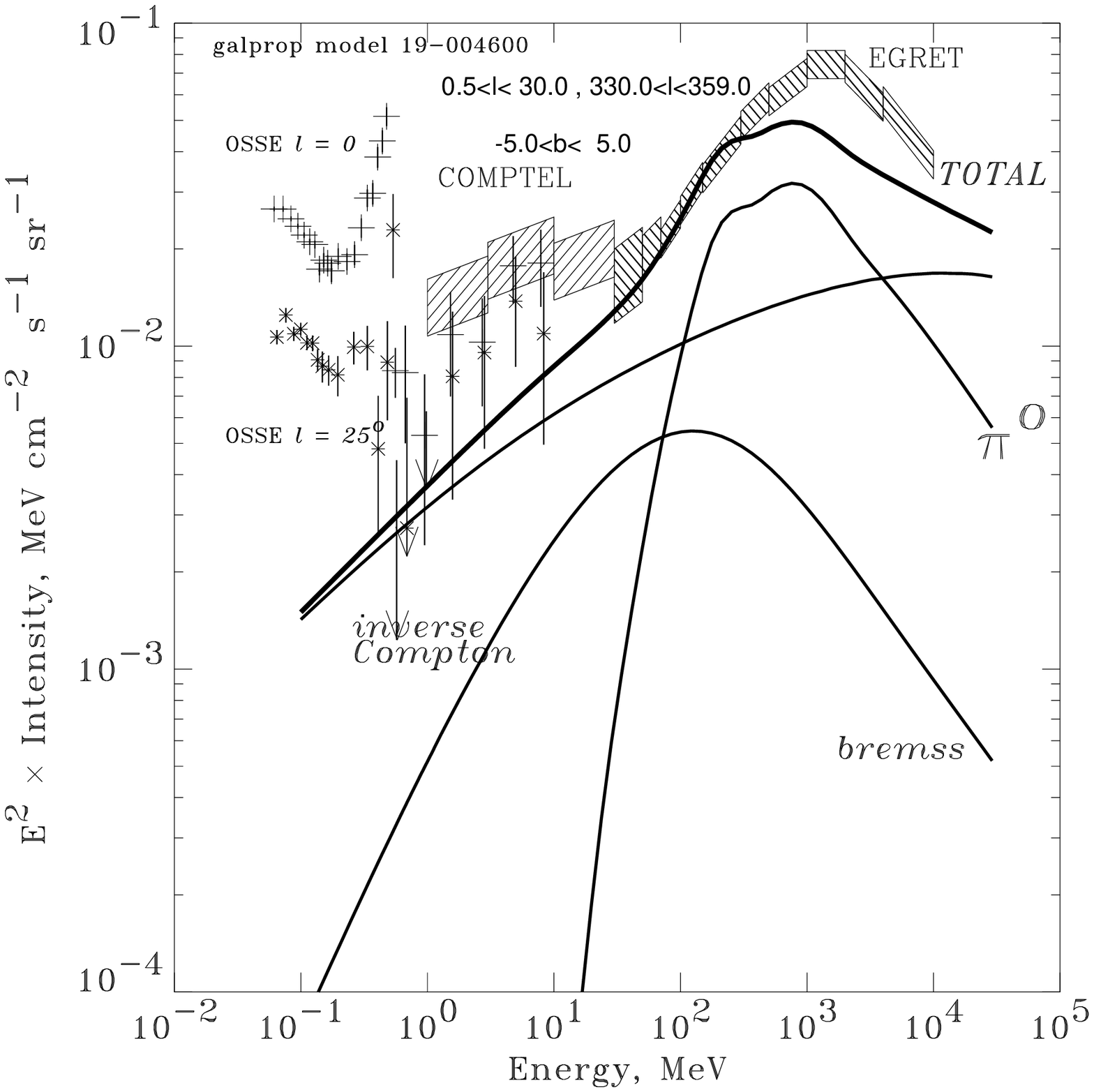,height=2.8in,width=2.8in }
   \epsfig{file= 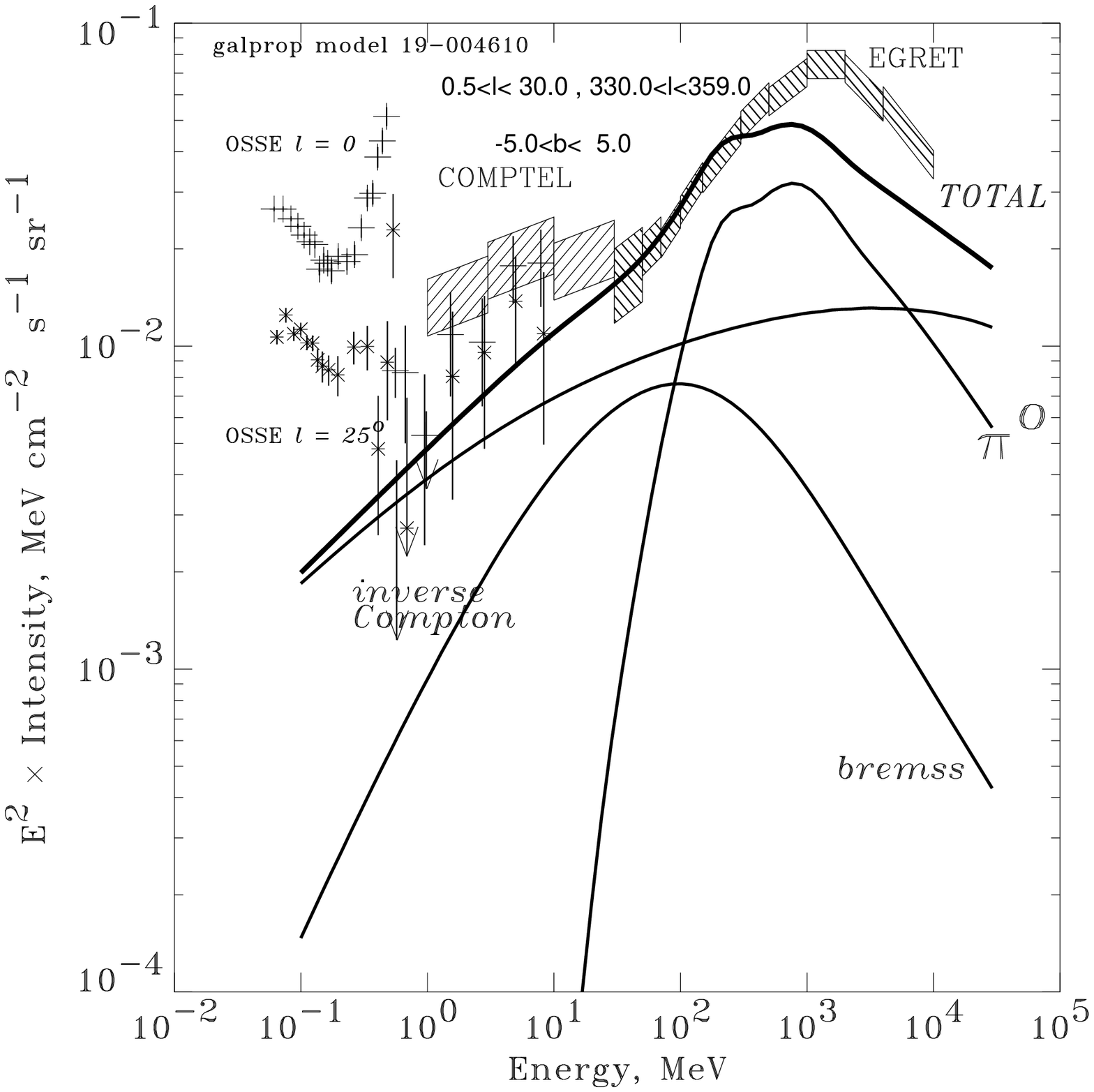,height=2.8in,width=2.8in }
   \epsfig{file= 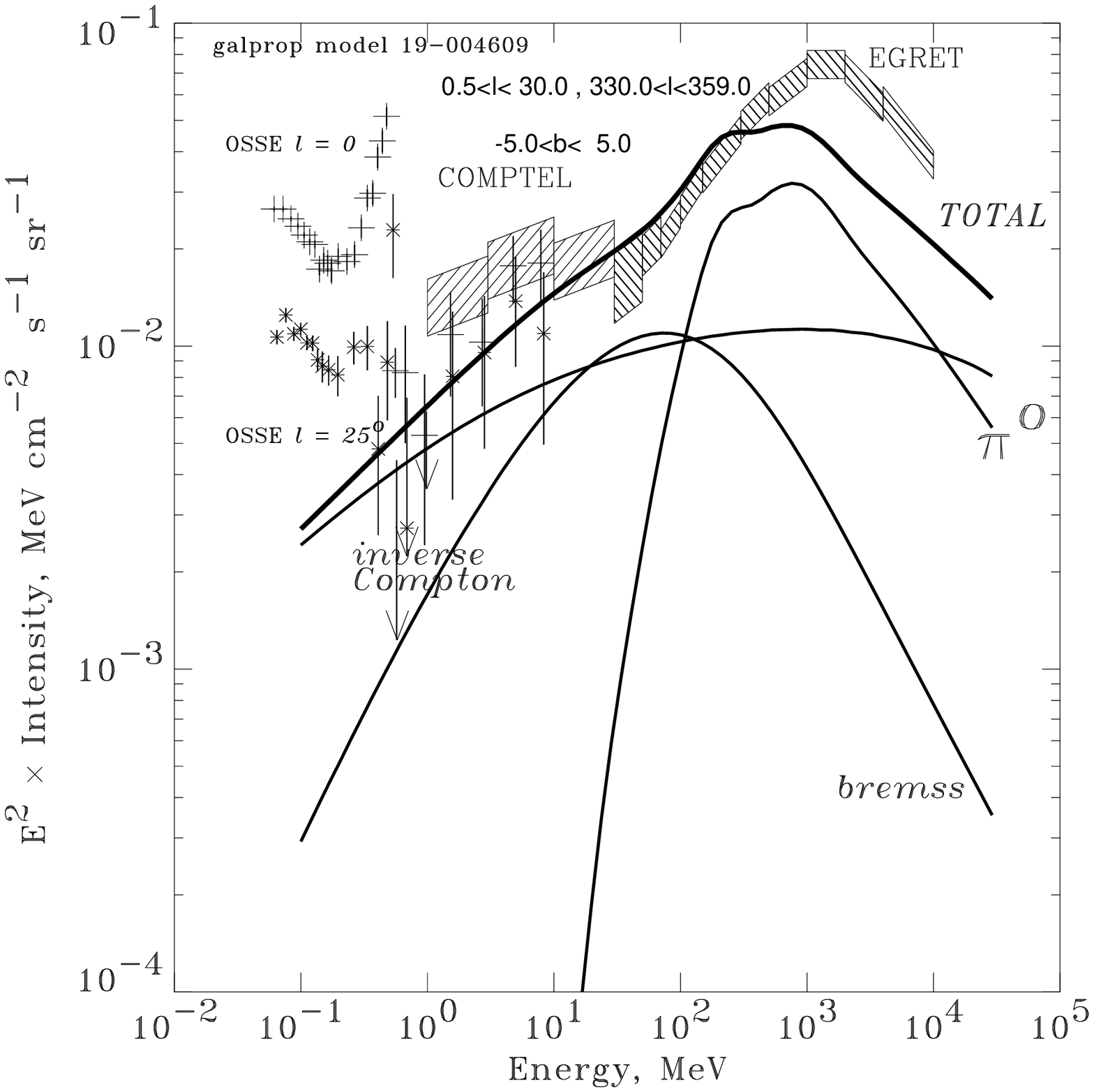,height=2.8in,width=2.8in }    
   \epsfig{file= 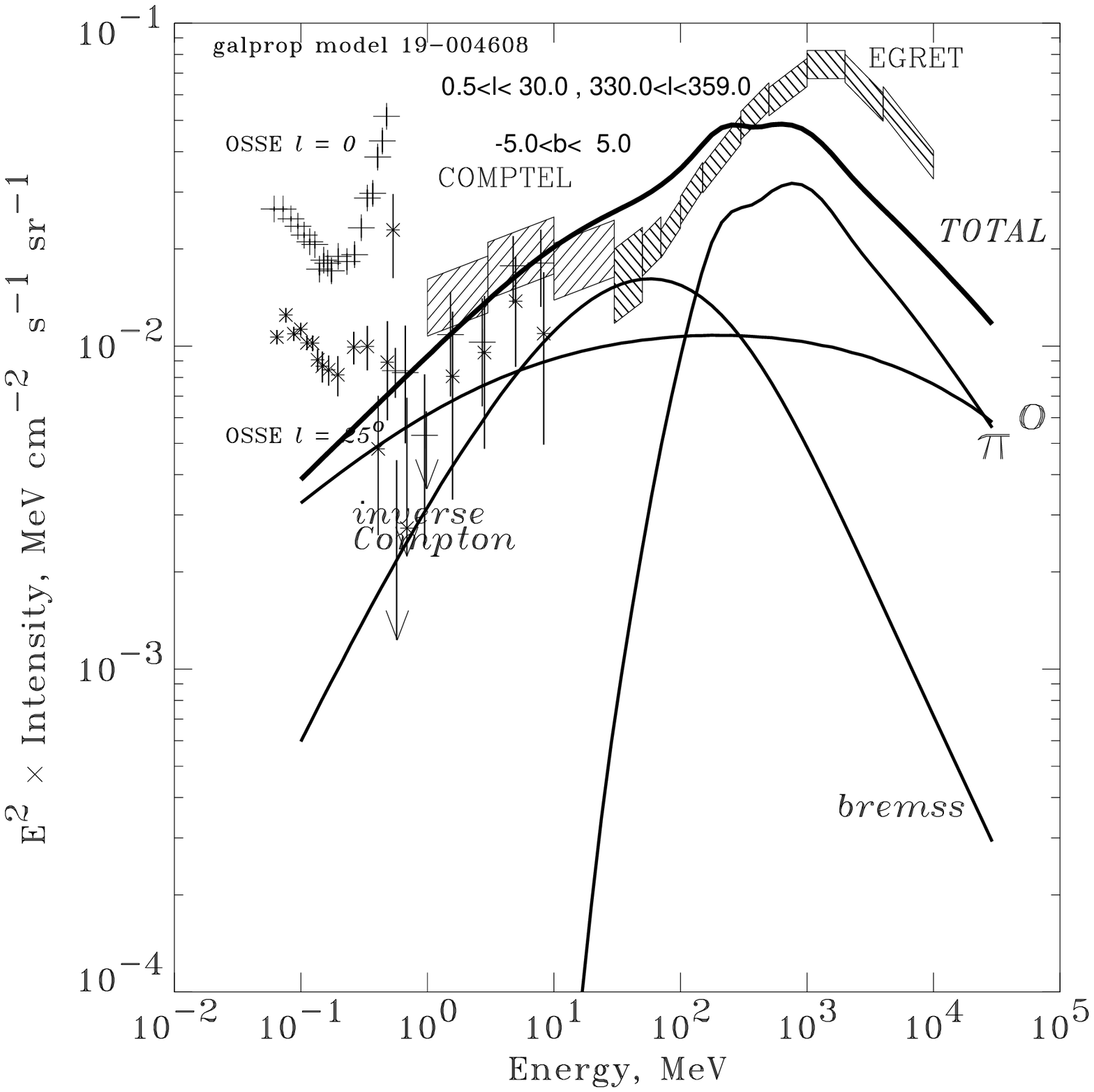,height=2.8in,width=2.8in }
   \centerline{
   \epsfig{file= 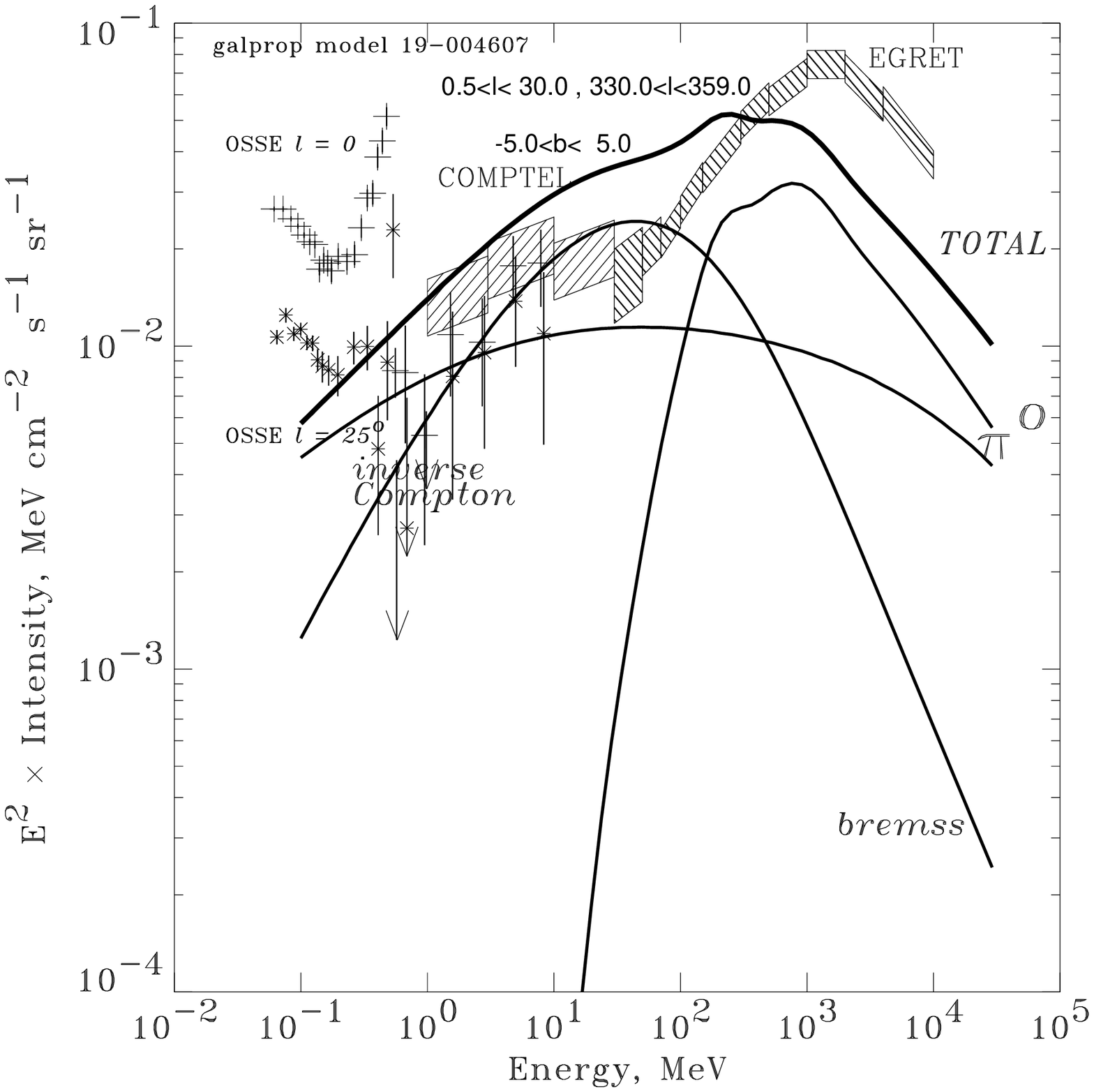,height=2.8in,width=2.8in }
   } 
\vspace{10pt}
\caption{Gamma-ray spectrum of the inner Galaxy for various electron
injection  spectra. Injection index 2.0 to 2.4 (from top to bottom),
no reacceleration.  Data:
\protect\cite{Kinzer,Strong98,StrongMattox96} (for details see
\protect\cite{smr98}).
}
\label{gamma_spectra}
\end{figure}

\begin{figure}[ht] 
   \centerline{\epsfig{file=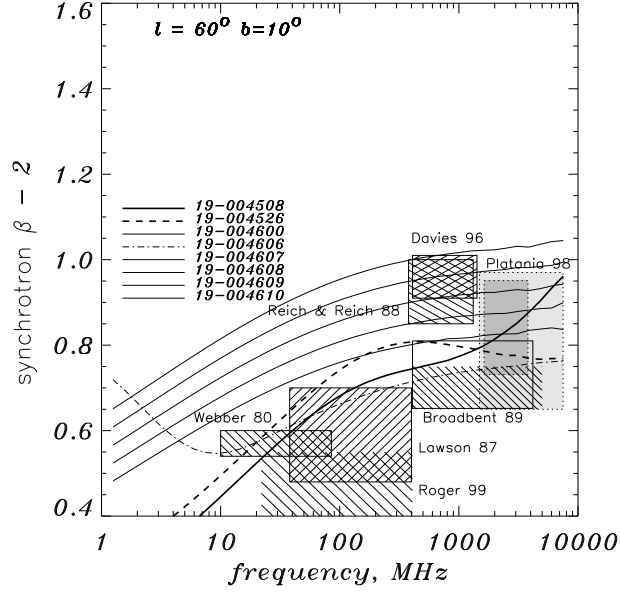,height=3.4in,width=3.5in}}
\vspace{10pt}
\caption{Synchrotron index for the electron injection spectra shown in
Fig \protect\ref{electrons}.  Thin solid lines (from bottom to top):
injection index 2.0 to 2.4.  Data:
\protect\cite{Roger,Lawson,Broadbent,Davies,Platania,Reich,Webber}
(for details see \protect\cite{smr98}).  
}
\label{synchrotron}
\end{figure}

\begin{figure} 
   \epsfig{file=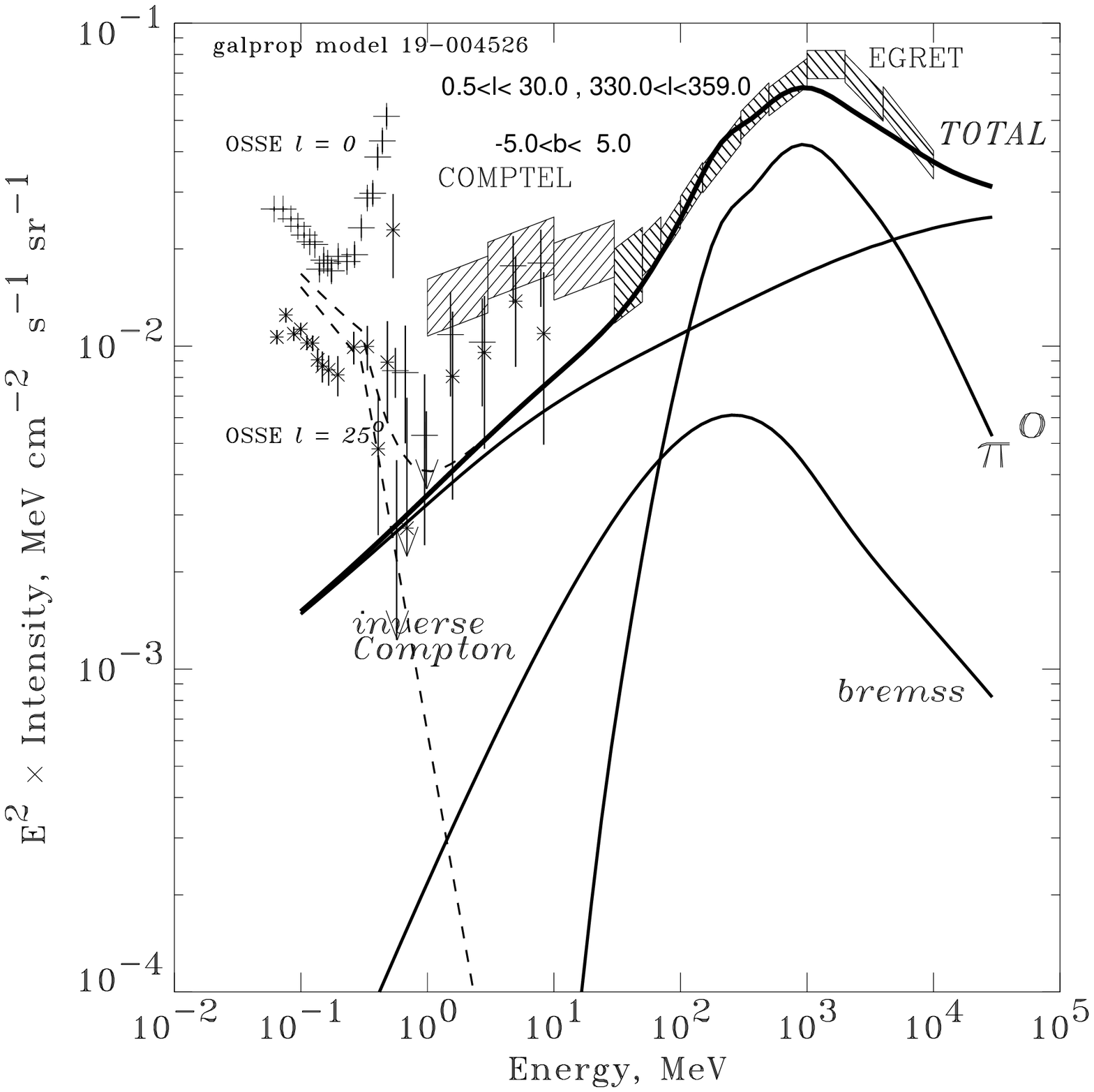,height=2.8in,width=2.8in}
   \epsfig{file=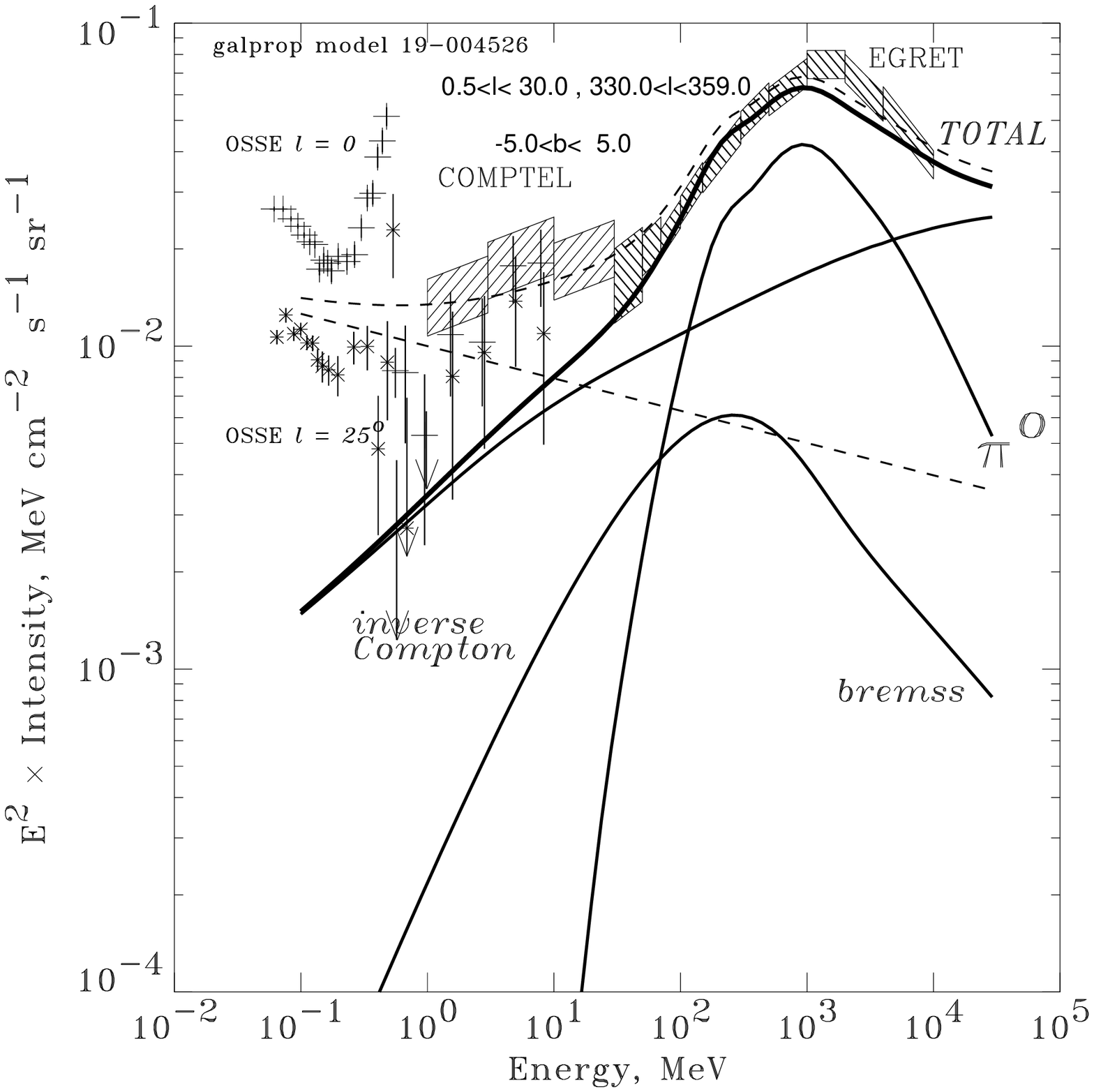,height=2.8in,width=2.8in}
\vspace{10pt}
\caption{Gamma-ray spectrum of the inner Galaxy with possible
unresolved source population components.  The dashed lines show the
assumed source contribution and the sum of source and diffuse
components.  The solid lines are the diffuse components alone.  Left:
Cyg X1- (low soft X-ray state)  like source spectrum; Right:
Crab-like source spectrum.  The electron injection spectrum is hard
(see text).  Data as Fig \protect\ref{gamma_spectra}.
}
\label{spectra_plus_sources} 
\end{figure}

\section{Electrons, $\gamma$-rays and synchrotron}

Conventionally the low-energy $\gamma$-ray continuum spectrum has been
explained by invoking a soft electron injection spectrum with index
2.1--2.4, and this could reproduce the 1--30 MeV emission as
bremsstrahlung plus inverse Compton emission (see e.g.\
\cite{Strong97}).  Fig \ref{electrons} shows a range of electron
spectra which result from propagation of injection spectral indices
2.0--2.4; the model is from \cite{smr98}; in order to illustrate more
clearly the effect these spectra are without reacceleration.  The
nucleon spectrum is consistent with local observations and is
described in \cite{smr98}.  Fig \ref{gamma_spectra} shows the inner
Galaxy $\gamma$-ray spectrum for the same electron spectra.  The best
fit is evidently obtained for index 2.2--2.3.

A problem with this, which was noted earlier but has become clearer
with more refined analyses, is the constraint from the observed
Galactic synchrotron spectrum on the electron spectral index above 100
MeV.  The synchrotron index is hard to measure because of baseline
effects and thermal emission, but there has been a lot of new work in
this area, in part because of interest in the cosmic microwave
background.  Fig \ref{synchrotron} summarizes relevant measurements of
the synchrotron index together with the predictions for the range of
electron spectra in Fig \ref{electrons}.  The new 22--408 MHz value
from \cite{Roger} is of particular importance here; it is consistent
with that derived earlier in a detailed synchrotron modelling study
\cite{Lawson}.  The $\gamma$-rays fit best for an injection index
2.2--2.3, but the synchrotron index for 100--1000 MHz is then about
0.8 which is above the measured range.  Although we illustrate this
for just one family of spectra for a particular set of propagation
parameters, it is clear that it covers the possible range of plausible
spectra so that changing the propagation model would not alter the
conclusion.  Hence we are unable to find an electron spectrum which
reproduces the $\gamma$-rays without violating the synchrotron
constraints.  If there were a very sharp upturn in the electron
injection spectrum below 200 MeV, as illustrated in Fig
\ref{electrons}, then we could explain the $\gamma$-rays as
bremsstrahlung emission without violating the synchrotron constraints,
but even then it would not reproduce the intensities below 1 MeV
measured by OSSE \cite{Kinzer}.

\section{An unresolved source population ?}

In view of the problems with diffuse emission we suggest that an
important component (at least 50\%) of the $\gamma$-ray emission below
10 MeV originates in a population of unresolved point sources; it is
clear that these must anyway dominate eventually as we go down in
energy from $\gamma$-rays to hard X-rays (see e.g.  \cite{Valinia98}),
so we propose the changeover occurs at MeV energies.  For illustration
we have tried adding (with arbitrary normalization) to the diffuse
emission possible spectra for the unresolved population (Fig
\ref{spectra_plus_sources}):  a low-state  Cyg X-1  type
\cite{McConnell2000} appears too steep, but a Crab-like type
($E^{-2.1}$) would be satisfactory, and would require a few dozen
Crab-like sources in the inner Galaxy.  These would not be detectable
as individual sources by COMPTEL and such a model not violate any
observational constraints which we know of.  In the examples in Fig
\ref{spectra_plus_sources} we have used the hard electron injection
spectrum (index 1.8) required to fit the $>$1 GeV excess
\cite{smr98,Strong2000} so that with Crab-like sources we can finally
reproduce the entire spectrum from 100 keV to 10 GeV.

This hypothesis has many  observational consequences which can only be
investigated by detailed modelling of source populations.


\begin{references}
\bibitem{sm98} 
   Strong, A.W., and Moskalenko I.V., {\it ApJ}\ {\bf 509}, 212 (1998).
   
\bibitem{smr98} 
   Strong, A.W., Moskalenko I.V., and Reimer, O., 
     {\it}\ {\bf} astro-ph/9811296 (1998).
   
\bibitem{Strong2000} 
   Strong, A.W., Moskalenko I.V., and Reimer, O., {\it these proceedings}.
   
\bibitem{Strong98}
   Strong, A.W. et al., {\it Proc 3rd INTEGRAL Workshop}, in press,
      astro-ph/9811211.
   
\bibitem{Bloemen2000}
   Bloemen, H., et al. {\it these proceedings}.

\bibitem{Strong97}
   Strong, A.W., and Moskalenko I.V., {\it AIP Conf. Proceedings}\ {\bf
     410}, 1162 (1997).

\bibitem{Roger}
   Roger, R.S., et al., {\it A\&AS}\ {\bf 137}, 7 (1999).

\bibitem{Lawson}
   Lawson, K.D., et al., {\it MNRAS}\ {\bf 225}, 307 (1987).

\bibitem{Kinzer}
   Kinzer, R.L., et al., {\it ApJ}\  {\bf 515}\  215, (1999).

\bibitem{Valinia98} 
   Valinia, A., and Marshall, F.E., {\it ApJ}\ {\bf 505}, 134 (1998).

\bibitem{McConnell2000}
   McConnell, M., et al., {\it these proceedings}.

\bibitem{StrongMattox96} 
   Strong, A.W.,  and Mattox, J. R.,  {\it A\&A }\ {\bf 308}, L21 (1996).

\bibitem{Broadbent}
   Broadbent, A., Haslam, C.T.G, and Osborne, J.L., {\it MNRAS}\ {\bf 237}, 
     381 (1989).

\bibitem{Davies}
   Davies, R.D., et al., {\it MNRAS}\ {\bf 278}, 925 (1996).

\bibitem{Platania}
   Platania, P., et al.l {\it ApJ}\ {\bf 505}, 473 (1998).

\bibitem{Reich}
   Reich, P., and Reich, W., {\it A\&A}\ {\bf 196}, 211  (1988).

\bibitem{Webber} 
   Webber, W.R., Simpson, G.A., and Cane, H.V., {\it ApJ}\ {\bf 236} 448 
     (1980).

\end{references}
\end{document}